\documentclass{pasj00}
\usepackage{rotating, epsfig}

\begin{document}

\Received{$\langle$15-05-09$\rangle$}
\Accepted{$\langle$20-03-10$\rangle$}

\author{Gwenifer Raymond$^{*1}$, Kate G. Isaak$^{1}$, Dave Clements$^{2}$, Adam Rykala$^{1}$, Chris Pearson$^{3,4}$}

\affil{
$^{*}$ gwenifer.raymond@astro.cf.ac.uk

$^{1}$ Cardiff University, School of Physics \& Astronomy, Queens
Buildings, The Parade, Cardiff, CF24 3AA, UK

$^{2}$ Astrophysics Group, Blackett Laboratory, Imperial College,
Prince Consort Road, London, SW7 2BW, UK

$^{3}$ Space Science and Technology Department, CCLRC Rutherford Appleton
Laboratory, Didcot, Oxfordshire OX11 0QX, UK

$^{4}$ Department of Physics, University of Lethbridge, 4401 University Drive,
Lethbridge, Alberta T1J 1B1, Canada 
} 

\title{The Effectiveness of Mid IR / Far IR Blind, Wide Area,
Spectral Surveys in Breaking the Confusion Limit} \date{24-11-09}

\maketitle
\begin{abstract}
\noindent 
Source confusion defines a practical depth to which to take large-area
extragalactic surveys. 3D imaging spectrometers with positional as
well as spectral information, however, potentially provide a means by
which to use line emission to break the traditional confusion
limit. In this paper we present the results of our investigation into
the effectiveness of mid/far infrared, wide-area spectroscopic surveys
in breaking the confusion limit. We use SAFARI, a FIR imaging Fourier
Transform Spectrometer concept for the proposed JAXA-led SPICA
mission, as a test case.  We generate artificial skies representative
of 100 SAFARI footprints and use a fully-automated redshift
determination method to retrieve redshifts for both spatially and
spectrally confused sources for bright-end and burst mode galaxy
evolution models.  We find we are able to retrieve accurate redshifts
for 38/54\% of the brightest spectrally confused sources, with
continuum fluxes as much as an order of magnitude below the 120 $\mu$m
photometric confusion limit.  In addition we also recover accurate
redshifts for 38/29\% of the second brightest spectrally confused
sources.  Our results suggest that deep, spectral line surveys with
SAFARI can break the traditional photometric confusion limit, and will
also not only resolve, but provide redshifts for, a large number of
previously inaccessible galaxies.  To conclude we discuss some of the
limitations of the technique, as well as further work.
\end{abstract} 

                                                                              
\section{Introduction}
The cosmic infrared background (CIB) peaks at $\sim 150$ $\mu$m and
comprises the total infrared (IR) emission from all sources in the sky
(eg. \cite{dole-01}, \cite{elbaz-02}), integrated over all time.  It
has been found to contain as much energy as the combined optical/UV
extragalactic background, suggesting that half of all light emitted by
stars and active galactic nuclei (AGN) is absorbed by dust before we
are able to observe it in the optical \citep{hauser-dwek}.  Locally,
the IR output of typical galaxies is only one third of their optical
output \citep{soifer-neugebauer-91}, which implies strong evolution in
the IR properties of galaxies as one moves to high redshift. The CIB
has been well-studied using many instruments including ISOCAM (ISO),
MIPS (Spitzer), ISOPHOT (ISO) and SCUBA(JCMT) at 15, 24, 160 and 850
$\mu$m (\cite{elbaz-02}, \cite{papovich-04}, \cite{juvela-00} and
\cite{smail-02} respectively).  Although the peak of the CIB lies at
$\sim150$ $\mu$m, it has yet to be resolved into individual sources at
these wavelengths: our understanding of the make-up of the CIB
therefore relies on extrapolation.

Observations in the mid and far-infrared (MIR and FIR; typically
defined as lying in the wavebands 5-30 and 30-1000 $\mu$m
respectively) from ground based telescopes are difficult if not
impossible because of the high opacity of the Earth's atmosphere at
these wavelengths.  Ground based telescopes are only sensitive to
narrow wavebands in the MIR/FIR where the atmosphere's transmission is
higher, thus wide band observations in the MIR/FIR must typically be
made from space.  As a result, FIR telescopes are smaller in diameter
than their optical counterparts (up to a few meters), with a resulting
angular resolution that is low compared to both optical and radio
telescopes/interferometers. A direct consequence of this low angular
resolution is that completely resolving the CIB into discrete sources
in the FIR is very difficult, if not impossible, because of source
confusion.

Source confusion may be defined as the degradation of the quality of
photometry of sources clustered on a scale to the order of the
telescope beam size (eg. \cite{scheuer-57}).  The confusion limit sets
the useful depth to which large-area extra-galactic surveys should be
taken.  For example, the Herschel mission \citep{pilbratt-04},
successfully launched in May 2009, has a 3.5 m diameter mirror which
realizes an angular resolution of 8'' at 120 $\mu$m. At these
wavelengths, the confusion limit for such a mirror is estimated to be
around $\sim$5 mJy (eg. \cite{dole-04}, \cite{jeong-06}).  Surveys at
24 $\mu$m suggest that at these flux levels one will only be able to
resolve at most $\sim$~50\% of the CIB \citep{dole-04}.  It is
possible to reduce the confusion limit through making observations
with a larger diameter mirror.  However, due to the practical
limitations of high angular resolution FIR imaging there is a limit on
how much we can reduce confusion noise.

One way to break the confusion limit makes use of the extra dimension
of wavelength, to which one has access in spectroscopic surveys.
Discrete sources can be identified by relatively bright, narrow-band
emission lines: thereby allowing redshifts to be determined.  A
preliminary study to explore the efficacy of blind, wide area
spectroscopic surveys in resolving FIR sources is described in
\citet{clements-07}.  In their work an artificial `sky' was populated
using template FIR spectral energy distributions (SEDs) of a selection
of different types of galaxy, to which were added FIR emission lines
of strengths derived from ISO-LWS observations
(eg. \cite{negishi-01}).  The sources were redshifted and assigned
luminosities according to the evolutionary models of
\citet{pearson-05}, \citet{pearson-07} and
\citet{pearson-k-09}. Observations of the `sky' were made using the
instrumental parameters (eg. sensitivity/noise levels, spectral
resolution, field of view (FoV), beam size) of SAFARI, a FIR imaging
Fourier Transform Spectrometer concept for the proposed JAXA-led SPICA
(Space Infrared Telescope for Astronomy and
Astrophysics) \citep{swinyard-08} mission.  It will offer the large FoV
and high spectral resolution required to break the confusion limit
using spectroscopy.  According to its current specifications SPICA
will have a 3.5 m diameter mirror, and therefore will be subject to
the same confusion noise as Herschel.  The primary mirror will,
however, be cooled to $<$6 K and so will offer a great leap in
sensitivity over Herschel.  SAFARI will cover the waveband 35 to 210
$\mu$m with varying resolution, including R$\sim$1000 - (at 120
$\mu$m, $\Delta\lambda = 0.176$ $\mu$m, when run in SAFARI's higher
resolution mode) - which matches the typical width of an extragalactic
MIR/FIR emission line.

Estimates of source redshift were made by hand by locating the
position of the strongest emission line in each spectrum.  The
strongest lines typically observed in the FIR are the [OI] and [CII]
lines at 63.18 and 157.74 $\mu$m respectively. If one assumes a typical
dust temperature of 35 K, then the strongest lines shortward and
longward of the peak of the SED will be [OI] and [CII] respectively.
Beyond $z=2.5$ these lines are shifted out of SAFARI's observable
waveband, therefore the redshifts of more distant sources than this
are irretrievable.  If these two lines were the only ones present in
the FIR then by comparing the evaluated source redshifts with the
model input redshifts, one can assess the efficiency of this
blind-line method.  It was found that when looking at a patch of
simulated `sky' equal to one SAFARI FoV, it was possible to retrieve
accurate redshifts for sources with 120 $\mu$m continuum fluxes as
much as a factor of $\sim$10 below the traditional continuum confusion
limit.  Sources with 120 $\mu$m flux $S_{120\mu m} > 1$ mJy and at
redshifts $z < 2.5$ were retrieved with 100\% accuracy.

The use of blind spectral line surveys to resolve FIR sources is not
without its own limitations and type of confusion. Line, or spectral
confusion occurs when multiple sources are observed in a single
telescope beam: the spectra from two or more objects are effectively
scrambled, and it can become difficult to determine which lines are
emitted by which objects. As a result, source redshifts become hard to
extract.  The work described in \citet{clements-07} made use of model
spectra with FIR emission lines only. To assess the true viability of
using spectral line surveys to break the confusion limit requires the
inclusion of MIR emission lines in the `sky' model, as sources will, in
general, have both FIR and MIR emission lines.  Inclusion of these
shorter wavelength lines will enable the recovery of sources with
redshifts of $z > 2.5$, beyond which the [OI] and [CII] emission lines
at 63.18 and 157.74 $\mu$m, respectively, are shifted out of the SAFARI
waveband.  By including MIR lines however, one increases the problem
of line confusion, and so assigning lines to individual, but spatially
unresolved, sources becomes more problematic.

In this paper we examine a much larger model `sky' populated with more
realistic template spectra with both FIR and MIR emission lines and
employ a new automated method of evaluating source redshifts in a time
efficient manner.  We also implement a method to extract the redshifts
of multiple sources clustered in a single spatial bin.  Through the
implementation of this method we investigate how effectively we can
break the traditional photometric confusion limit.  In
sections~\ref{sec:generate_sky} and~\ref{sec:generate_cube} we
describe the generation of the artificial sky used to test the source
recovery technique, which in turn is outlined in
section~\ref{sec:method}. This is followed in
section~\ref{sec:results} by a quantitative assessment of the
retrieval and error rates of the redshifts output, and a discussion of
the results in section~\ref{sec:discussion}.  

\section{Extracting Redshifts From an Artificial Sky}

Deep, blind-field imaging spectroscopy has the potential to enable
discrete sources at fluxes below the traditional continuum confusion
limit to be extracted.  Using imaging spectroscopy, it is
theoretically possible to extract redshifts for all line emitting
sources present in the instrument's FoV, allowing all sources to be
discretely resolved.  In this work this is done using an automated
redshift determination algorithm.  In order to best test the
efficiency (number of sources for which we determine redshifts, inversely
weighted by how many sources we inaccurately determine redshifts for)
of this method we generate an artificial sky in the form of a
datacube, populated with realistic spectra taken from nearby galaxies,
and redshifted according to the bright-end and burst mode galaxy
evolution models of \citet{pearson-05}, \citet{pearson-07} and
\citet{pearson-k-09}.  Running the program through the datacube we
compare the fitted redshifts with their input values in order to
determine the method's precision, accuracy and efficiency.
  
\subsection{Generating an Artificial Sky}
\label{sec:generate_sky}

We create an initial set of two 1 square degree artificial skies with
1'' spatial pixels, populated by galaxies with redshifts, spectral
types and 40 $\mu$m continuum fluxes based on the bright-end and burst
mode evolutionary models of \citet{pearson-05}, \citet{pearson-07} and
\citet{pearson-k-09}.  We do not include any physically based spatial
distribution modelling (eg. clustering, etc.) and in this work the
skies are populated with sources uniformly distributed in random
positions.  It should also be noted that we do not include any cirrus
contribution and we assume SAFARI to be background limited.  These act
as our master 'skies' which we later crop and re-bin to create the
SAFARI footprints.

The bright-end and burst mode evolutionary models are backward
evolution formulations where observed galaxy source counts are used to
constrain the model parameters. The model components consist of a
luminosity function to represent the number density of sources as a
function of luminosity, a library of spectral energy distributions
(SED) to model the extragalactic source population emission as a
function of wavelength and an assumption on the type-dependent
evolution of the extragalactic population (in luminosity and number
density).  Both evolutionary models utilize the IRAS infrared local
luminosity function defined at 60$\mu$m \citep{saunders-00} or
12$\mu$m \citep{rush-93} for the galaxy and AGN (Seyfert) populations
respectively. Although various other, more recent luminosity functions
are available, the IRAS functions has the advantage of being defined
at or around the peak of the population emission spectrum and are free
of contamination by mid-infrared features.  The 60$\mu$m galaxy
luminosity function is segregated into cool (normal galaxy) and warm
(star-forming) components, defined by {\it IRAS} colours where cool
100 $\mu$m/60$\mu$m cirrus-like colours \citep{efstathiou-rr-03}
represent the normal quiescent galaxy population and the warmer 100
$\mu$m/60 $\mu$m colour component is representative of star-forming
galaxies with activity increasing as a function of luminosity for
M82-like Starburst $L_{IR}<10^{11}L_{\odot}$, luminous (LIRG)
$L_{IR}>10^{11}L_{\odot}$ \& ultraluminous (ULIRG)
$L_{IR}>10^{12}L_{\odot}$ infrared galaxies. Thus the model framework
includes five general evolutionary population classes (Normal,
Starburst, LIRG, ULIRG, AGN) of extragalactic object defined by
luminosity, colour and subsequent evolution.

The bright-end model assumes evolution in the galaxy population in
both the density and luminosity for sources modeled by simple power
laws of the form $f(z) =(1+z)^{k}$, where $k$ is the type dependent
evolutionary strength parameter.  The evolutionary model is an updated
framework of that first presented in \citet{pearson-rr-1996}, with
modest starburst galaxies rather than ULIRGs dominating in 15 $\mu$m
selected source counts.  The burst mode evolutionary model predicts
that the upturn of emission at 15 $\mu$m \citep{elbaz-99} and peak at
24 $\mu$m \citep{papovich-04} is to due the emergence of a new
population of L/ULIRGs.  The original burst mode evolutionary model
presented by \citet{pearson-2001} has power law evolution similar to
the bright-end model for the starburst and AGN sources and an initial
violent exponential evolutionary phase of the form, $f(z) = 1 + f
.exp[-{(z-z_{p})^{2}\over{2\sigma ^2}}]$ , from $z=0$ to $z_{p}=1$,
where where $k$ \& $\sigma$ are the type dependent evolutionary
strength parameters, followed by a power law evolutionary phase for
the L/UILRGs.  Both the bright-end and burst mode evolutionary models
have non-evolving normal galaxy populations.  Throughout this work,
values regarding the different evolution models are written in the
form bright-end(burst mode).

Each of the five general evolutionary population components (Normal,
Starburst, LIRG, ULIRG, AGN) are represented by a small set of galaxy
spectral energy distributions from the libraries of
\citet{efstathiou-rr-03}, \citet{efstathiou-00} and
\citet{efstathiou-rr-95} for the normal, starburst, L/ULIRG and AGN
types respectively.  The selected SEDs are representative of the SED
libraries from which they have been drawn and have been shown to be
consistent with the colours of sources detected in the European Large
Area ISO Survey \citep{oliver-2000}.  \citet{rowan-robinson-04} showed
that the ISO infrared galaxy population could indeed be divided into
four general spectral classes; normal quiescent galaxies; starburst
(M82-like) galaxies; luminous and ultraluminous infrared galaxies and
AGN.  \citet{rowan-robinson-04} found that the normal, quiescent
population of ISO galaxies were well modelled with the templates of
\citet{efstathiou-rr-03} with far/mid-infrared ratios of $\nu
S_{\nu}$(100$\mu$m)/$\nu S_{\nu}$(12$\mu$m)$\sim$6-7 whilst in
contrast, IRAS galaxies were often modelled with templates with ratios
of $\sim$5 \citep{rowanrobinson-89}.  Therefore, the normal galaxy
component consist of 2 SEDs each of $\nu S_{\nu}$(100$\mu$m)/$\nu
S_{\nu}$(12$\mu$m)$\sim$5.8 \& 6 referred to as a normal and cold
normal type respectively.  For the star forming (starburst, L/ULIRG)
population we have selected 7 SEDs from the template libraries to
ensure a variation in the the mid-infrared features in an attempt to
avoid artifacts caused by a particular choice of SED for all
sources. The SEDs in the template libraries have a broad correlation
between the model optical depth and the galaxy luminosity.  2/7 of
these SEDs are selected for the starburst component ($\tau_{V} \sim
50$ of which one is a model for the archetypal star-forming galaxy
M82). 3/7 of these SEDs are selected with increasing optical depths
$(\tau_{V} \sim 50-100$) for the LIRG component (assuming a
corresponding increase in luminosity for each SED of $10^{11}
L_{\odot}$, $10^{11.5} L_{\odot}$ and a colder SED of $L>10^{11.5}
L_{\odot}$ referred to as a cold-LIRG) and the remaining 2/7 SEDs
selected for the ULIRG component correspond to the best template model
fits for the archetypal ULIRGs Arp220 (cold ULIRG) and Mk231 (hot
ULIRG) respectively.  Given the relatively featureless infrared
spectra of AGN, the AGN component SED corresponds to a single tapered
disc dust torus model.  Emission lines are then added to each of the
template spectra, taken from ISO-LWS observations of nearby galaxies
(eg. \cite{negishi-01}).

\begin{figure*}[htbp]
  \centering
  \epsfig{figure=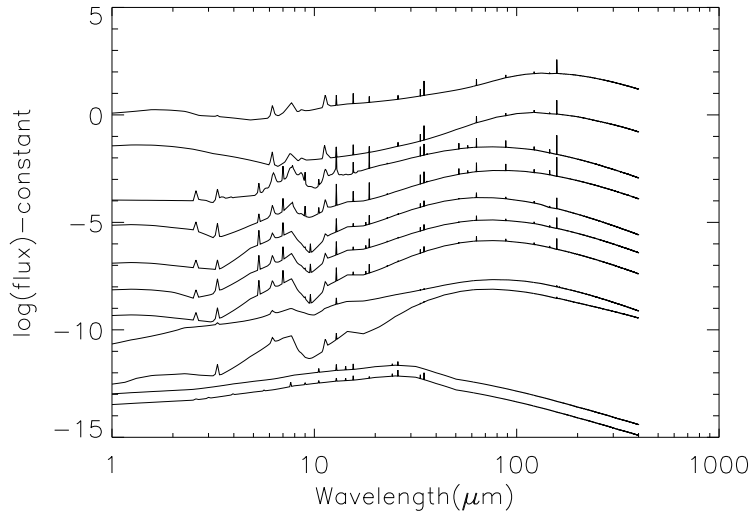, height = 3.0in}
  \caption{Template spectra which were used to populate the datacube.
  From the top downwards are the SED templates for; normal cold,
  normal, starburst M82, starburst, $10^{11}L_{\odot}$ LIRG,
  $10^{11.5}$ $L_{\odot}$ LIRG, cold LIRG, hot ULIRG, cold ULIRG,
  Seyfert 1 and Seyfert 2.}
  \label{fig:templates}
\end{figure*}

MIR emission line strengths are taken, where possible, from the same
sources as are used for the FIR emission lines.  In some cases data
were not currently available at these shorter wavelengths in which case
sources with similar FIR characteristics as the original template
spectra are used.  The galaxies from which we take the MIR emission
line strengths are listed in table~\ref{mirtable}.  The model spectra
are shown in figure~\ref{fig:templates} and are added to the master
`sky' in spectral resolution of $\Delta\lambda = 0.08$ $\mu$m (R =
2500 at 200 $\mu$m) in the waveband from 1 to 400 $\mu$m.

\begin{table*}[htbp]
  \centering
  \begin{tabular}{cccc}
    \hline
    Component & Galaxy Type & Line Template & Reference \\
    \hline
    Normal & Cold/Normal & NGC 7331 & \cite{smith-04} \\
    Starburst & M82/Starburst & M82 & \cite{forsterschreiber-01} \\
    LIRG & $10^{11/11.5} L_{\odot}$/Cold LIRG & NGC 253 & \cite{sturm-00} \\
    ULIRG & Hot/Cold ULIRG & Arp 220 & \cite{sturm-96}  \\
    AGN & Seyfert 1 & Mrk 1014 & \cite{armus-04} \\
    AGN & Seyfert 2 & NGC 1068 & \cite{lutz-00} \\
  \end{tabular}
  \caption{Sources used for the addition of MIR lines to the template spectra that populate the data cube.}
  \label{mirtable}
\end{table*}

\begin{figure*}[h!tbp]
	\centering \epsfig{figure=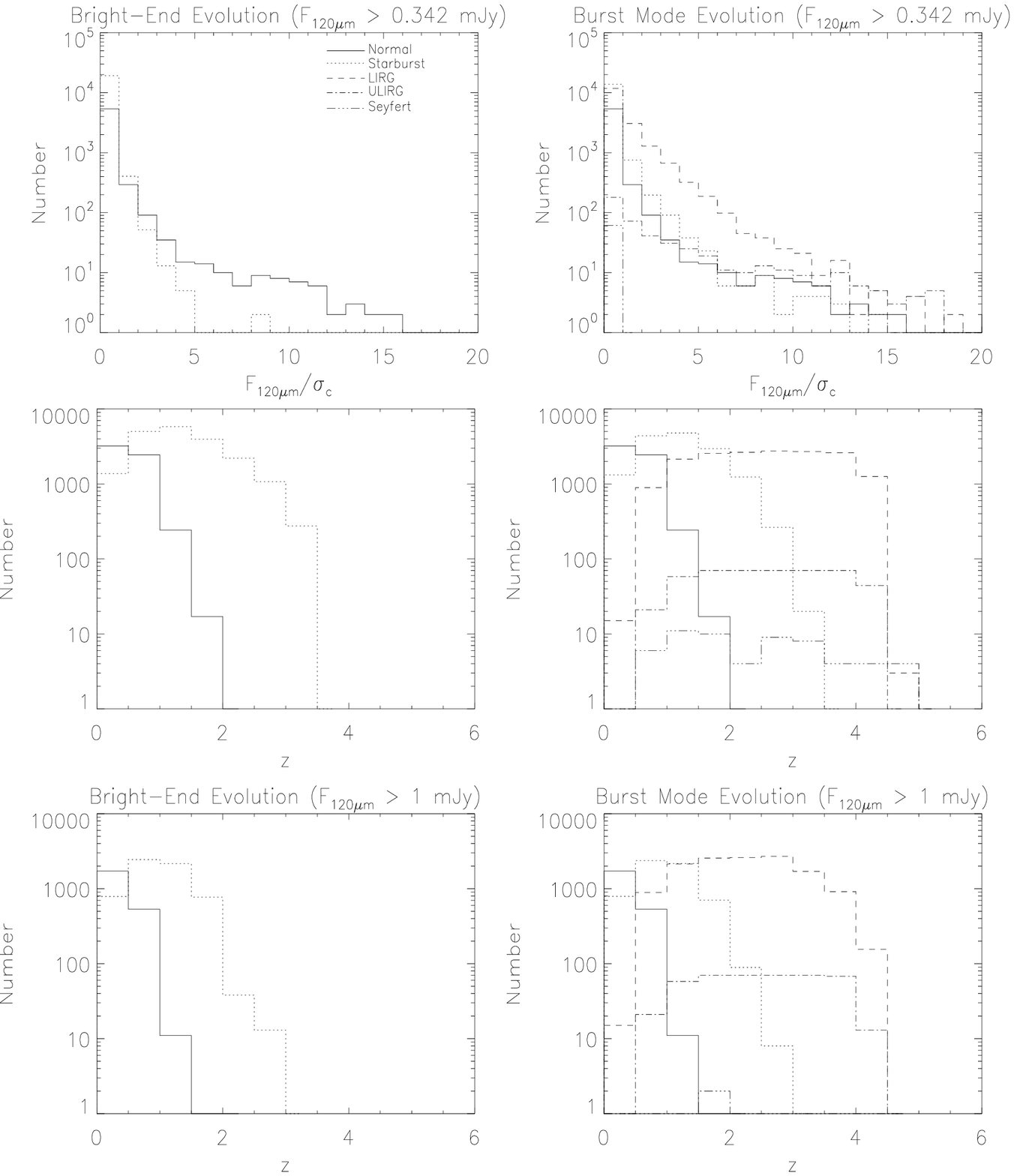, height = 7.5in}
		\caption{A plot of the flux (top row) and redshift
		(bottom two rows) distributions, for each SED type, of
		the sources that populate the bright-end and burst
		mode evolution data cubes respectively.  The top two
		rows are for sources with 120 $\mu$m flux, $S_{120\mu
		m} > 0.342$ mJy and the bottom row is for sources with
		$S_{120\mu m} > 1$ mJy.  Values are for a 1 square
		degree region of sky.  Flux is measured as a fraction
		of the 120 $\mu$m confusion limit, $\sigma_{c}=4.3$
		mJy}
		\label{fig:models}
\end{figure*}

The flux and redshift distributions for each SED type for the
bright-end and burst mode evolution models are shown in
figure~\ref{fig:models}.  These figures show the distribution of SED
type with 120 $\mu$m flux and redshift over a 1 square degree region
of sky.  In these regions of sky there are a total of 25596(38975)
sources with $S_{120\mu m} > 0.342$ mJy (1$\sigma$, 10hr sensitivity
of SAFARI).  Of these sources 33(58)\% have $S_{120\mu m} > 1$ mJy.
The burst mode evolution model includes more high flux sources as well
as more high redshift sources than the bright-end model.  The burst
mode model also includes the more extreme luminous/ultraluminous and
Seyfert spectral types, which tend to dominate at higher redshifts.
100(75)\% of sources with $S_{120\mu m} > 1$ mJy have redshifts $z <
2.5$, however setting $S_{120\mu m} > 0.342$ mJy these values drop to
95(74)\%.

\subsection{Generating Datacubes/Source Catalogs}
\label{sec:generate_cube}

SAFARI \citep{swinyard-08} will have a 2'$\times$2' FoV with a
diffraction-limited angular resolution of 8'' at 120 $\mu$m.  The
instrument will cover the waveband from 35 to 210 $\mu$m at varying
spectral resolution (R$\sim$1000 at 120 $\mu$m).  However, in this
work we match our data cubes to the original specifications of SAFARI,
taking a FoV of 128''$\times$128'' and a waveband covering from 30 to
210 $\mu$m.  Thus to generate datacubes representative of the sky as
would be seen by SAFARI we take 128''$\times$128'' sections of our
artificial skies and re-bin to a spatial resolution of 8'' (to
simplify we have assumed a wavelength independent spatial and spectral
resolution).  This re-binning allows for the possibility of two or
more sources in a single spatial bin.  In these cases we refer to the
brightest source with $S_{120\mu m} > 0.342$ mJy as the primary source
and the second brightest $S_{120\mu m} > 0.342$ mJy as the secondary
source.  Applying the angular resolution of SPICA means that 6(8)\% of
all pixels have two or more sources present.  If there are multiple
sources present in a spatial pixel only the two brightest will be
investigated as other sources will be too faint relative to the
primary and secondary sources to detect.  We create our spectra by
cropping the waveband of the artificial sky to between 30 and 210
$\mu$m with and smoothing them to a fixed resolution of $\Delta\lambda
= 0.176$ $\mu$m.  In this work we use the original wavelength range
SAFARI rather than its current specification, thus we have a slightly
larger waveband to pick up emission lines from.  We then add Gaussian
noise with a standard deviation of $\sigma = 0.342$ mJy and a zero
mean along each spectrum\footnote{The is the noise commensurate with
the sensitivity of SAFARI for a 10hr integration time.}.  A datacube
representing a single SAFARI footprint will therefore be 16x16x1024
pixels in size.

A `truth' catalog is simultaneously generated which tracks the
location of each source in a datacube, as well as the redshift, SED
type and 120 $\mu$m flux.  This allows a later comparison between
retrieved and input redshifts after we apply our redshift
determination algorithm.  This is the same datacube and `truth'
catalog generation method as used by \citet{clements-07}.

Datacubes are generated for both burst mode and bright-end
evolutionary models.  For each evolutionary model 100 different
datacubes are investigated (i.e. taken from different regions of the
larger 1 square degree cube) in order to account for variance in the
random spatial distributions, each of a size equivalent to 1 SAFARI
FoV (i.e. a total area of 2'$\times$2').  For each cube the noise
along the spectra is added by creating 10 differently seeded randomly
generated Gaussian noise arrays in order to account for the variance
of results due to noise. We therefore have a total of 1000 datacubes
for each evolutionary model.  By way of comparison, a figure of
$\sigma_{c}(\lambda = 120$ $\mu m) = 4.3$ mJy is adopted for the
confusion limit, based on \citet{dole-04}.

\subsection{Detecting Sources and Extracting Their Redshifts}
\label{sec:method}
\begin{figure*}[h!tbp]
  \centering
  \epsfig{figure=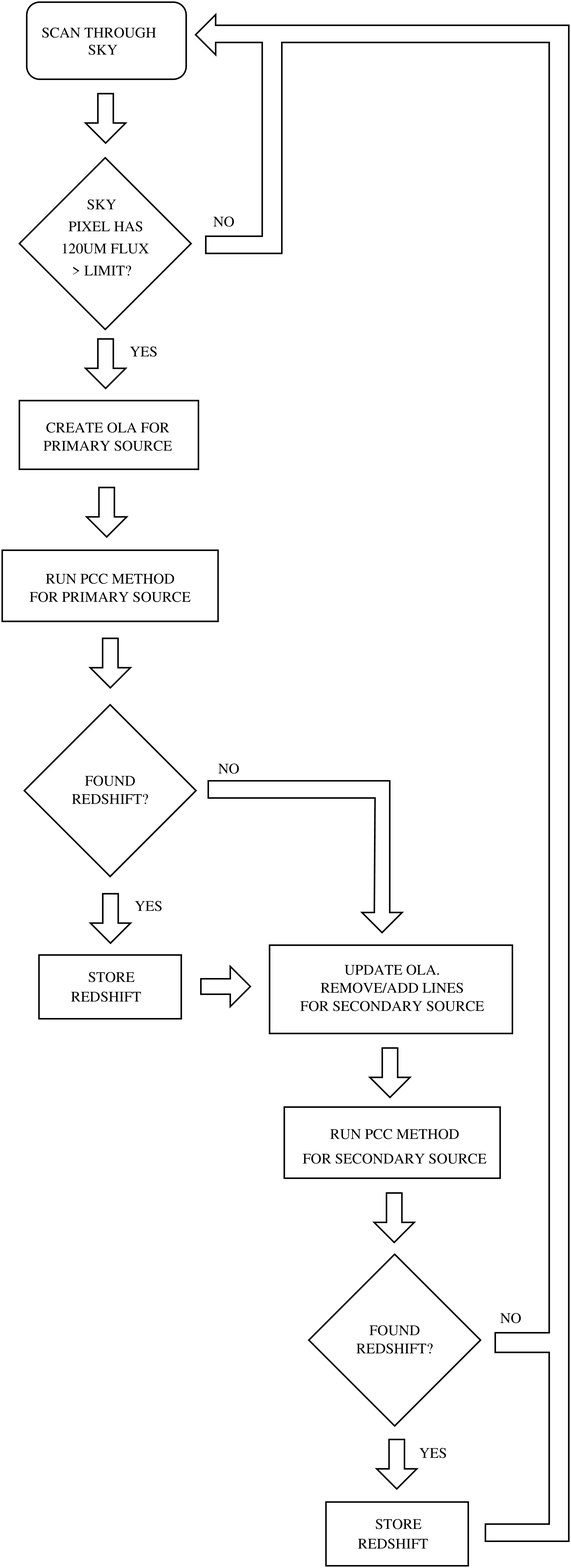, height = 9.0in}
  \caption{A flowchart of the order in which the sources are
  observed and then our redshift determination method implemented.}
  \label{fig:flowchart}
\end{figure*}

Redshifts are determined using a pseudo-cross-correlation (PCC) method
and stored - however, first the presence of a source must be
confirmed.  This is done by checking the 120 $\mu$m flux, $S_{120\mu
m}$, of each spatial pixel in the cube against a limiting value.  This
action is performed before any other operations take place, thereby
saving processing time on analyzing non-existent or too-faint sources.
We assume that if any spatial pixel has $S_{120\mu m} > $0.342 mJy
(which is equivalent to the 1$\sigma$ noise in a single spectral pixel
for 10 hours of integration) then a source is present and we attempt
to determine its redshift.  This value was chosen as our continuum
cutoff value as empirically it is found that if a source has 120
$\mu$m flux less than this then typically the emission lines are too
faint to reliably use with our method.  A flow diagram illustrating
the sequence of this method is shown in figure~\ref{fig:flowchart}.

\begin{figure*}[h!tbp]
  \centering \epsfig{figure=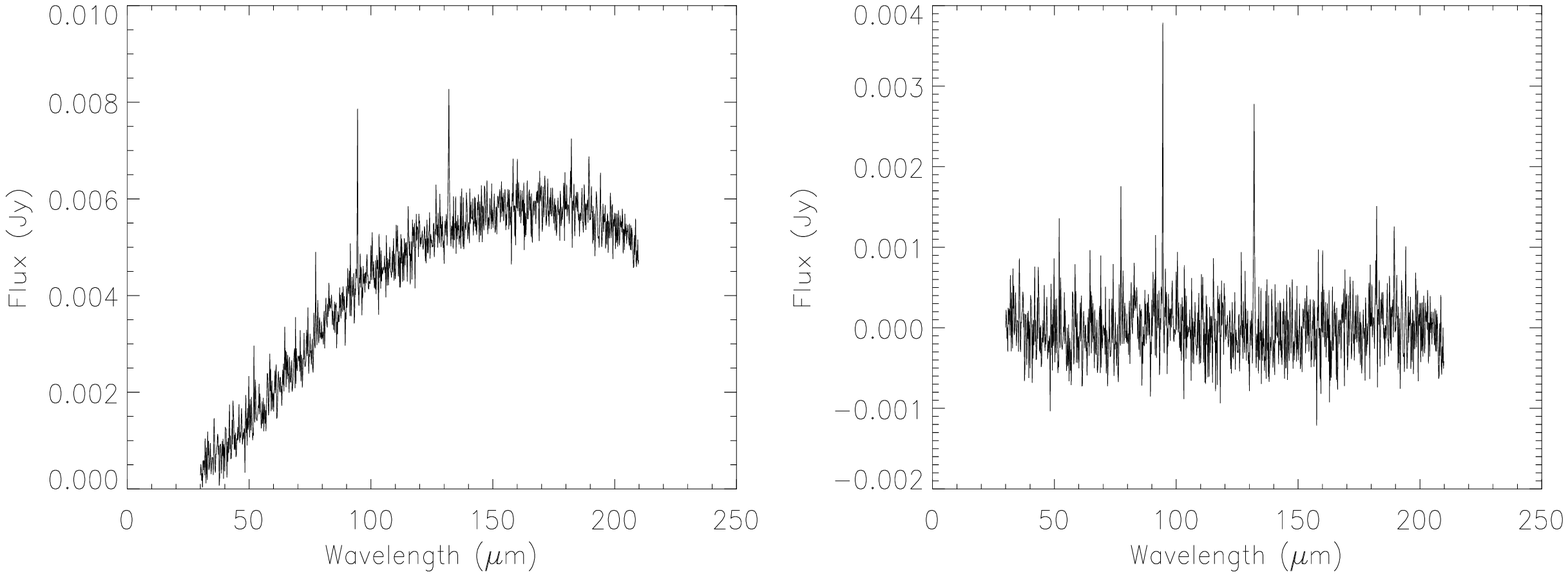, height = 2.5in}
  \caption{Shown in the left panel is a sample spectrum as can be
    found in our artificial `skies'.  In our PCC redshift
    determination algorithm the continuum of this spectrum is fit with
    a fourth order polynomial.  This fit is then subtracted from the
    spectrum, leaving an array containing only emission lines and
    noise.  The right panel shows the example spectrum after the
    continuum fit has been subtracted.}
  \label{fig:spec}
\end{figure*}

Each spectrum is preprocessed prior to redshift determination in the
following way: a) Each spectrum is fit with a fourth order polynomial
which is taken to represent the continuum emission of the source; b)
the $S_{120\mu m}$ value of the spectrum is checked against a limiting
value: if $S_{120\mu m} <$ the limiting value, the source is
considered too faint to determine its redshift, and no further
analysis is conducted on the spectrum; c) The polynomial continuum fit
is subtracted from the spectrum, leaving an array containing only
emission lines and noise (see figure~\ref{fig:spec})

The spectral channels in the aforementioned array containing the 6
highest continuum subtracted flux levels (with $S_{\lambda}>2\sigma$)
are then initially considered to be emission lines and are taken to be
our observed lines in our observed line array ($OLA$).  Empirically it
is found that if the ratio of the strongest to weakest (continuum
subtracted) line fluxes present in $OLA$ is less than 1.5 then these
lines most likely arise from noise and we therefore consider such an
array to contain no genuine emission lines.  Further analysis is only
conducted on spectra with higher ratio values than this.

$OLA$ is 1024 spectral channels in extent (i.e. identical to the
spectral extent of our data cube), and contains the
continuum-subtracted flux of our 6 observed possible emission lines at
the appropriate spectral channels corresponding to the wavelengths of
those lines, and zero elsewhere.  In the case of an emission line
stretching across multiple pixels, the line is compressed into a
single channel and there is a built in tolerance in the redshift
fitting routine to allow for this.  The brightest channel present in
$OLA$ is assumed to hold a genuine emission line.  The wavelength of
this emission line is then compared to the set of template wavelengths
compiled from a list of the strongest emission lines typically seen in
galactic spectra in the MIR/FIR.  This template line array (TLA) is
equal to one at the appropriate spectral channels corresponding to the
wavelengths of the template emission lines, and zero elsewhere.  This
comparison then provides a table of possible redshifts the source may
lie at.

If the strongest line in $OLA$ lies at $\lambda_{brightest}$, and our
template emission lines lie at wavelengths $\lambda_{template_{j}}$,
where $j$ is the indexing of the line in our template (eg. for
[OIII]@51.82 $\mu$m $j = 1$, for [NIII]@57.32 $\mu$m $j = 2$ etc.),
then an array of possible redshifts at which the source may lie can be
determined from:

\begin{equation}
z_{k} = \frac{\lambda_{brightest}}{\lambda_{template_{j}}} - 1
\end{equation}
Where $k=$ 1,2,\textellipsis$l$, where l is the total number of lines
in $TLA$ and thus the total number of possible redshifts.  The array
of template emission lines, $TLA$, which was used to determine the set
of possible redshifts is now also used to fit to the observed possible
emission lines, $OLA$.  Thus in order to find the redshift at which
$TLA$ most closely correlates to $OLA$, $TLA$ is shifted to each of
the possible source redshifts, as in;
\begin{equation}
\lambda_{template_{jk}} = \lambda_{template_{j}}(1+z_{k})
\end{equation} 
We now have $l$ binary redshifted template line arrays ($TLA_{k}$), each
of which are 1024 spectral channels in extent and are defined by;
\begin{equation}
   (TLA)_{ik}(\lambda_{i}=\lambda_{template_{jk}}) = 1 
\end{equation}
\begin{equation}
(TLA)_{ik}(\lambda_{i}\neq\lambda_{template_{jk}}) = 0
\end{equation}
where $i$ is the spectral channel corresponding to a given wavelength.
The strength of the correlation between our observed line array and
each template line array is given by;
\begin{equation}
C_{k} = \sum_{i=0}^{i = 1023} (OLA)_{i}(TLA)_{ik}
\end{equation}

The value of $z_{k}$ which gives the highest value for $C_{k}$ is then
assigned as the best estimate of the source redshift.  In this way the
strength of the match depends both on the strength of the continuum
subtracted emission lines that are coincident between the observed
spectrum and the redshifted line templates, and the number of matches
between lines of the observed and template emission line arrays.
 
To improve the accuracy of our method, use is made of additional,
a-priori information, and redshifts which produce template/observed
array emission line matches for strong, commonly observed emission
line pairs are weighted more heavily.  The additional weighting values
for $C_{k}$ when a characteristic pair is found are determined
empirically to give the most favorable ratio of accurate to
inaccurate source redshift evaluations\footnote{An evaluated source
redshift is defined as being accurate if it differs from the input
model redshift by less than 0.1, i.e. $|z_{evaluated}-z_{catalog}| =
\Delta z < 0.1$}
\footnote{Taking as an example three different weighting values: $A$,
$B$ and $C$.  $A$ outputs 3 accurate redshifts and 0 inaccurate
redshifts.  $B$ outputs 20 accurate redshifts and 10 inaccurate
redshifts.  $C$ outputs 15 accurate redshifts and 3 inaccurate
redshifts.  Of these weighting values we would use $C$ as this outputs
a high number of accurate redshifts, while limiting the number of
inaccurate redshifts.}.  Eg. one of the strongest line combinations in
the MIR is the [SIII]/[SiII] pair at rest wavelengths of 33.42 and
34.82 $\mu$m respectively, both of which are found to be strong in
starburst galaxies. If a match is made between the observed line array
and this line pair, then $C_{k}$ is weighted by an extra factor of 2.
Two other, weaker, line combinations included in our template array
are the [OIII]/[NIII] and [OIII]/[NII] pairs at rest wavelengths of
51.82, 57.32, 88.36 and 121.90 $\mu$m respectively.  If a match is
made between the observed line array and either of these line pairs,
then $C_{k}$ is weighted by an extra factor of 1.5.

We include an additional criterion that, as the strengths of the
members of the [OIII]/[NIII] and [OIII]/[NII] lines pairs are typically
comparable, any value of $z_{k}$ which gives a match in the observed
line array with one member of either of these pairs, but not its
partner, is rejected.  This relationship holds for all the model
spectra used in this work, however it may not always hold in practice
when encountering genuine spectra as observed in extragalactic
surveys.

In contrast to the work described in \citet{clements-07}, we can no
longer assume that the strongest emission line in the spectrum
shortward of the continuum emission peak (determined from the position
of the peak of the polynomial fit) is the [OI] line at 63.18 $\mu$m,
as we now have strong MIR lines present in the spectra.  We can,
however, still assume that the strongest line longward of the
continuum emission peak is the [CII] line at 157.74 $\mu$m.  This is
true as long as the reasonable assumption that $T_{dust} \simgtr 20$~K
holds, as [CII] is then typically the only strong line longward of the
SED peak.  Thus, if no single redshift is able to map the line
template onto the observed line array, but the strongest line in the
spectrum lies longward of the SED peak, and has a line to continuum
ratio $> 3$ (found empirically to be the lowest value to reliably use
to identify the [CII] emission line at 157.74 $\mu$m in this
circumstance), we assume it to be the [CII] line and, from this,
calculate a redshift.

An evaluated redshift is recorded along with the position of the
source on the sky.  The redshift is referred to as the primary
redshift and is considered to be that of the brightest, or primary,
source in the given spatial bin. A slightly modified redshift
extraction algorithm is then run a second-time through the spectrum,
to determine whether there is a second source of lower flux present;
the secondary source.  When attempting to extract a redshift for the
secondary source in any spatial bin we first zero the lines in $OLA$
which we have already associated with the primary source.  We then add
to $OLA$ the 3 strongest lines in the spectrum that have not yet been
used in redshift fitting to the spectrum, with flux $S_{\lambda} >
3.5\sigma$\footnote{Eg. defining a line as any spectral channel which
has $S_{\lambda} > 2\sigma$, if a spectrum contains 20 spectral
channels with $S_{\lambda} > 2\sigma$, we populate $OLA$ with the
brightest 6 of these lines, leaving 14 unused lines in the spectrum.
In order to determine a secondary redshift we zero the lines in the
$OLA$ which we have already associated with the primary source.  If,
for example, 4 lines in the $OLA$ contributed to the highest value of
$C_{k}$ then these are zeroed, leaving 2 lines remaining in $OLA$
which can be used to determine a secondary redshift.  Of the 14
remaining unused lines above $2\sigma$ in the spectrum, 10 of these
for example may have fluxes $S_{\lambda} > 3.5\sigma$, the 3 brightest
of which we add to $OLA$.  The $OLA$ used to attempt to determine a
secondary redshift therefore contains a total of 5 lines.}.  This line
selection process is used as a) We find empirically that 6 is the
optimum number of emission lines required to accurately fit a primary
redshift (i.e. the maximum amount of lines the algorithm is able to
use before encountering significant degeneracies) b) By definition the
secondary source is fainter than the primary, thus we expect noise to
be more of a significant hindrance in redshift fitting - therefore we
need more stringent requirements on the strength of the emission lines
used c) 3 is found to be the optimum number of lines to add to the
previously selected lines which are not found to be associated with
the primary source - if all 6 lines of the initially selected lines
are found to be associated with the primary source then 3 emission
lines is the minimum number that can be used to reliably fit a
redshift and any more than this can results in significant
degeneracies in redshift fitting.  The algorithm now runs in a manner
very similar to before, however (1) by definition we expect the
continuum-subtracted emission lines from the secondary source to be of
lower flux than those of the primary, and so the requirement on the
ratio of the strongest to the weakest emission line in the observed
array is dropped to 1.3, and (2) also by definition we are looking at
sources of fainter continuum flux, and thus are more susceptible to
picking up spurious emission lines: we therefore no longer weight more
heavily redshifts which give matches for characteristic MIR/FIR
emission line pairs, and no longer allow redshifts to be determined
from the single [CII] emission line.  Both these changes in the
algorithm for secondary as opposed to primary redshift determination
are implemented as empirically they are found to give the most
reliable results.

The PCC method works by moving through each spectrum in the cube and
cross-correlating MIR/FIR line templates at a discrete rather than
continuous set of redshifts.  This means that the number of discrete
redshifts fitted is far less than the number of redshift steps
required when fitting continuously over a given range.  This means we
are using less of the spectrum and thus are less likely to encounter a
pixel with a high noise level which could be miss-identified as an
emission line.  We are therefore able to use weaker emission lines for
redshift determination, which enables the algorithm to probe more
deeply into the noise.  This is particularly powerful when looking at
secondary sources which typically will have weaker line fluxes than
primary sources

 \subsubsection{FIR Emission Lines Only}
\label{sec:firmethod}
As a first test of our automated redshift determination we compare how
efficiently our method works in comparison to that described in
\citet{clements-07}.  Clements at al. made use of spectra containing
FIR emission lines only, therefore a direct comparison of this
original method, and our PCC method as described in
section~\ref{sec:method} can only be made when using exactly the same
spectra.  To do this we use a variant of our PCC method which uses FIR
emission lines only.  The lines are listed in table~\ref{ztempfir}.
In addition, only the [OIII]/[NIII] and [OIII]/[NII] pairs at
51.82/57.32 and 88.36/121.90 $\mu$m respectively are used to
additionally weight $C_{k}$, as these are the only characteristic
pairs which lie in the FIR waveband as defined by Clements et al.

\begin{table*}[h!tbp]
  \centering
  \begin{tabular}{cccccccccccc}
   \hline Emission Line & OIII & NIII & OI & OIII & NII & OI & CII \\
    Wavelength ($\mu$m) & 51.82 & 57.32 & 63.18 & 88.36 & 121.90 & 145.53 & 157.74 \\
    \hline
  \end{tabular}
  \caption{Lines used in source redshift determination through
  template fitting for the FIR emission line only method.}
  \label{ztempfir}
\end{table*}

\subsubsection{MIR and FIR Emission Lines}
\label{sec:mirmethod}
The general version of our method by default uses both MIR and FIR
lines in our template array, which are listed in table~\ref{ztempmir}.
The version of the method also allows us to make use of all of our
characteristic emission line pairs.

\begin{table*}[h!tbp]
  \centering
  \begin{tabular}{cccccccccccc}
   \hline Emission Line & NeII & SIII & SIII & SiII & OIII & NIII & OI & OIII & NII & OI & CII \\
    Wavelength ($\mu$m) & 12.81 & 18.71 & 33.42 & 34.82 & 51.82 & 57.32 & 63.18 & 88.36 & 121.90 & 145.53 & 157.74 \\
    \hline
  \end{tabular}
  \caption{Lines used in source redshift determination through
  template fitting for the method using both FIR and MIR emission
  lines.}
  \label{ztempmir}
\end{table*}

\section{Results}
\label{sec:results}
All results given in this section were determined by taking the
averaged values for each of 100 FoVs ($\times$ 10 differently seeded
noise arrays) for both bright-end and burst mode evolution.  When
quoting results from this work we give the number of accurate
redshifts retrieved as a percentage of the total number of sources
within a given flux and redshift range in the datacube.  However the
number of inaccurate redshifts are given as a percentage of the total
number of redshifts output by the algorithm under the same
constraints\footnote{Eg. a datacube has 100 sources within the
specified redshift and flux range and the algorithm outputs 50
redshifts; 40 accurate and 10 inaccurate.  The results statement would
then be given in the following form; we retrieve 40\% of sources in
our flux and redshift range accurately, with 20\% of redshifts output
by the algorithm under the same constraints being inaccurate.}.

\subsection{Results From Method of \citet{clements-07}}
The work described in \citet{clements-07} made use of spectra
containing FIR emission lines only, and results were only determined
for the burst mode evolution model. They found that all sources with
redshifts at $z \leq 2.5$ with $S_{120\mu m} \geq 1$ mJy could be
retrieved.  Redshifts higher than this could not be determined, as
beyond $z=2.5$ the [OI] and [CII] lines are redshifted out of the
SAFARI pass-band.  Lower flux sources, as much as $\sim$10 times fainter
than the traditional continuum confusion limit were also retrieved,
albeit with lower efficiency.

\subsection{Analysis Using FIR Emission Lines Only}
\label{sec:firresults}
\subsubsection{Primary Sources}
Using the FIR emission line only version of the method outlined in
section~\ref{sec:firmethod} we find that we
recover accurate redshifts for 85(46)\% of all primary sources with
$S_{120\mu m} \geq 1$ mJy, with 5(10)\% of all primary redshifts
output by the algorithm under the same constraints being inaccurate.
The aforementioned recovery values are given as a percentage of all
sources, including those at $z > 2.5$, however by definition this
method is unable to retrieve redshifts for sources at $z > 2.5$.
Beyond $z = 2.5$ the redshift recovery rate drops to zero for both
evolutionary models.  For sources with $S_{120\mu m} \geq 1$ mJy lying
at $z < 2.5$ the recovery of redshifts for primary sources is
$\sim$85(64)\%.  Under the same constraints \citet{clements-07}
retrieved accurate redshifts for 100\% of the sources.  However in
that work redshifts were determined manually for each source whereas
in this work they are determined automatically.  Unlike Clements et
al. we are unable to assess each spectrum on a case by case basis.
Therefore in attempting to minimize the number of incorrect redshifts
output by the algorithm we are forced to limit the maximum possible
efficiency of the accurate redshift recovery of the algorithm.

By dropping our 120 $\mu$m flux cutoff to $S_{120\mu m} \geq 0.342$
mJy, we find that we recover accurate redshifts for 36(33)\% of all
primary sources with $S_{120\mu m} \geq 0.342$ mJy, with 12(13)\% of
all redshifts output by the algorithm under the same constraints being
inaccurate.  Taking into account only sources at $z < 2.5$ our
accurate recovery becomes 45(39)\%.  Despite this being a lower
recovery percentage than for $S_{120\mu m} \geq 1$ mJy the majority of
sources have lower fluxes than this.  As such by dropping the cutoff
flux from 1 mJy to 0.342 mJy we have increased the total number of
accurate redshifts recovered by 22(15)\%.

\subsubsection{Secondary Sources}
\label{sec:firmult}
Using our FIR emission line only PCC method we retrieve accurate
redshifts for $\sim$30(27)\% of all secondary sources when using a
flux cutoff of $S_{120\mu m} > 0.342$ mJy with $\sim$6(4)\% of all
secondary redshifts output by the algorithm under the same constraints
being inaccurate.

\subsection{Extending Analysis to MIR}
\begin{figure*}[h!tbp]
  \centering
  \epsfig{figure=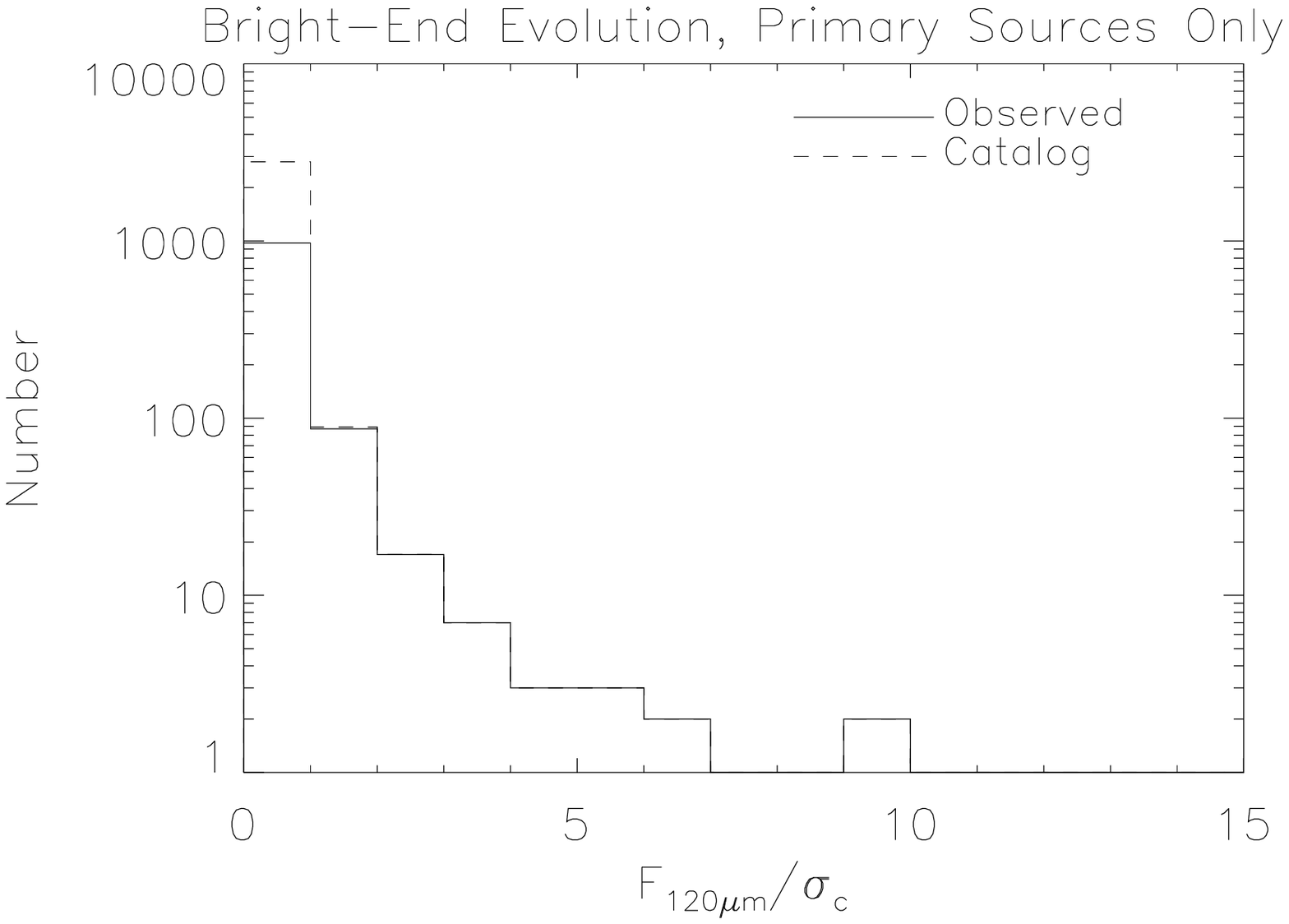, height = 2.5in}
  \epsfig{figure=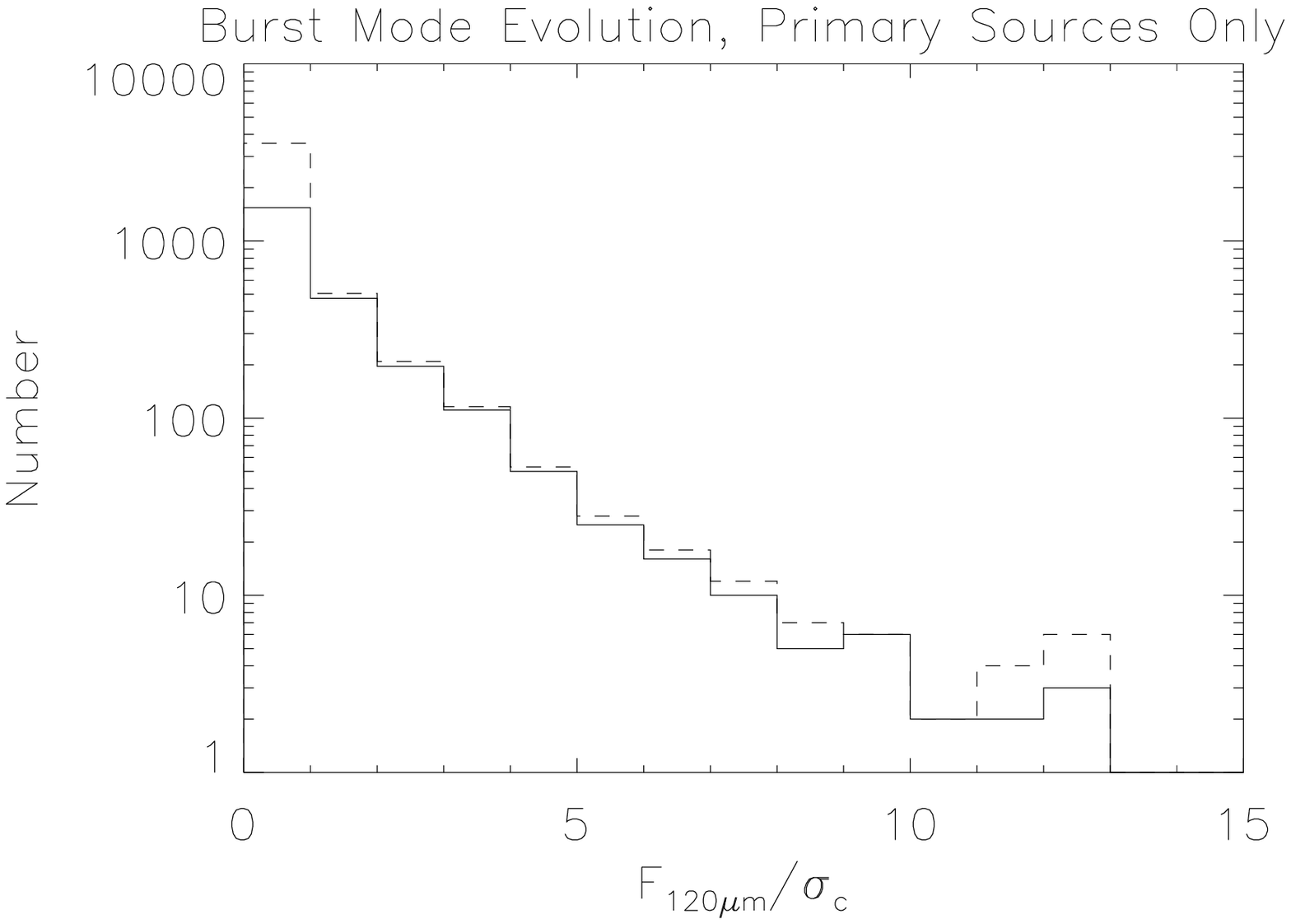, height = 2.5in}		
  \caption{Plots of the 120 $\mu$m continuum flux distribution (with
  flux measured as a fraction of the 120 $\mu$m continuum confusion
  limit, $\sigma_{c}(\lambda = 120\mu m) = 4.3$ mJy) for both the
  total input primary sources (dashed) and primary sources with
  accurately determined redshifts (solid).  Shown in the left and
  right hand panels are the results for the bright-end and burst mode
  evolution models respectively. Results are for 100 SAFARI FoVs.}
  \label{fig:flux1}
\end{figure*}

By default, the full version of our PCC redshift determination method
uses both MIR and FIR emission lines, and is that version which would
be applied to real data.  It is now possible to investigate sources at
redshifts $z > 2.5$ which account for 5(26)\% of the population with
$S_{120\mu m}>0.342$ mJy, of the evolutionary models.  Accurate
redshifts are retrieved for $\sim$75\% of primary sources with
$S_{120\mu m} \geq 1$ mJy, for both bright-end and burst mode
evolution, with respectively 6(8)\% of all primary redshifts output by
the algorithm under the same constraints being inaccurate.  We see
here that using both MIR and FIR emission lines is less efficient in
recovering accurate redshifts than using FIR emission lines only at
redshifts $z<2.5$.  This is due to the fact the ability to resolve
redshifts $z>2.5$ is not as advantageous here as most sources with
$S_{120\mu m} \geq 1$ mJy lie at redshifts $z<2.5$ in both evolution
models, however we are still subject to the disadvantage of
significantly more degeneracies in redshift fitting due to a higher
number of emission lines.  Dropping the 120 $\mu$m flux cutoff,
38(54)\% of all primary sources with $S_{120\mu m} \geq 0.342$ mJy are
retrieved (see figure~\ref{fig:flux1}), with 14(9)\% of all redshifts
output by the algorithm under the same constraints being inaccurate.

\subsubsection{Secondary Sources}
\begin{figure*}[h!tbp]
  \centering
  \epsfig{figure=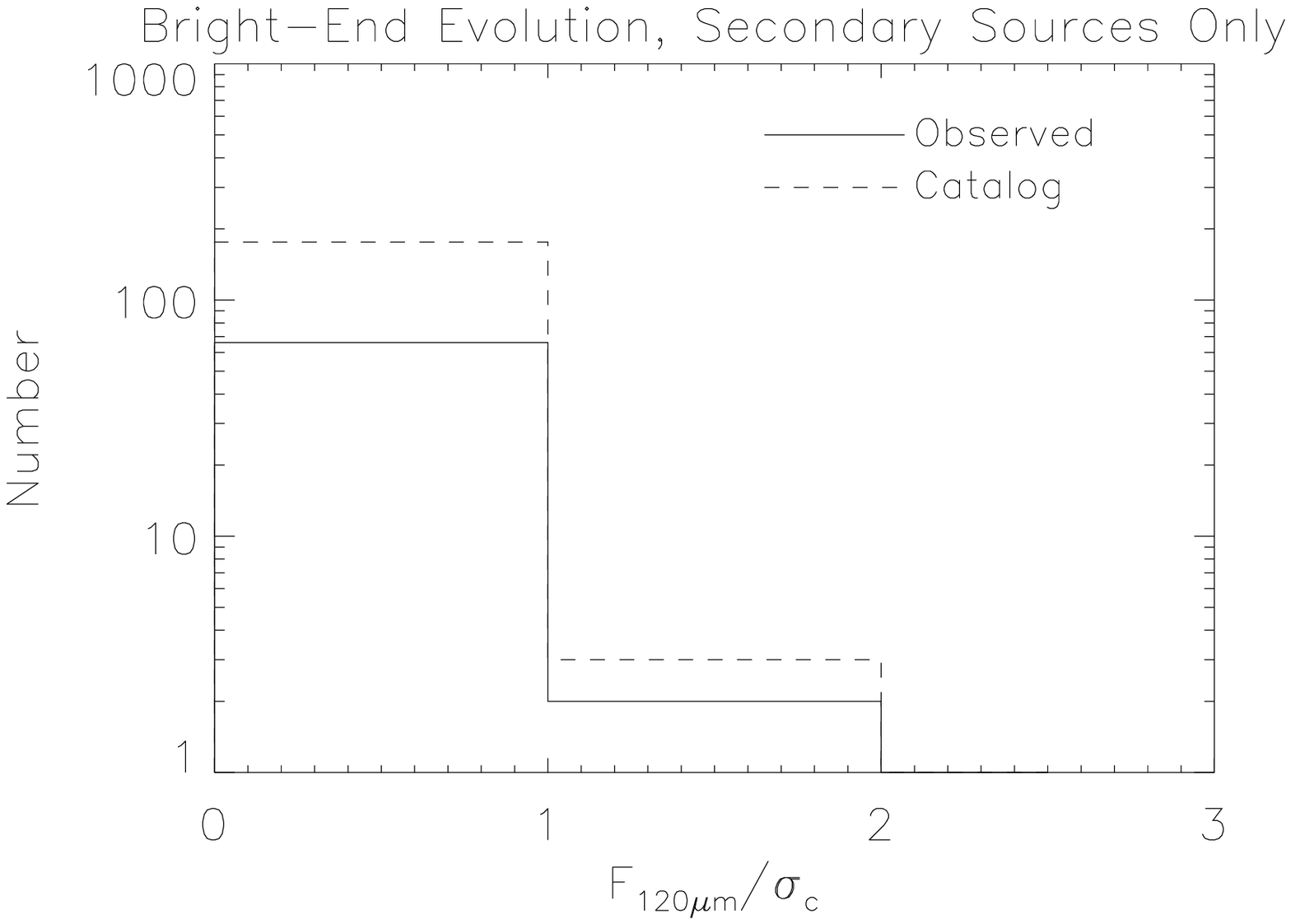, height = 2.5in}
  \epsfig{figure=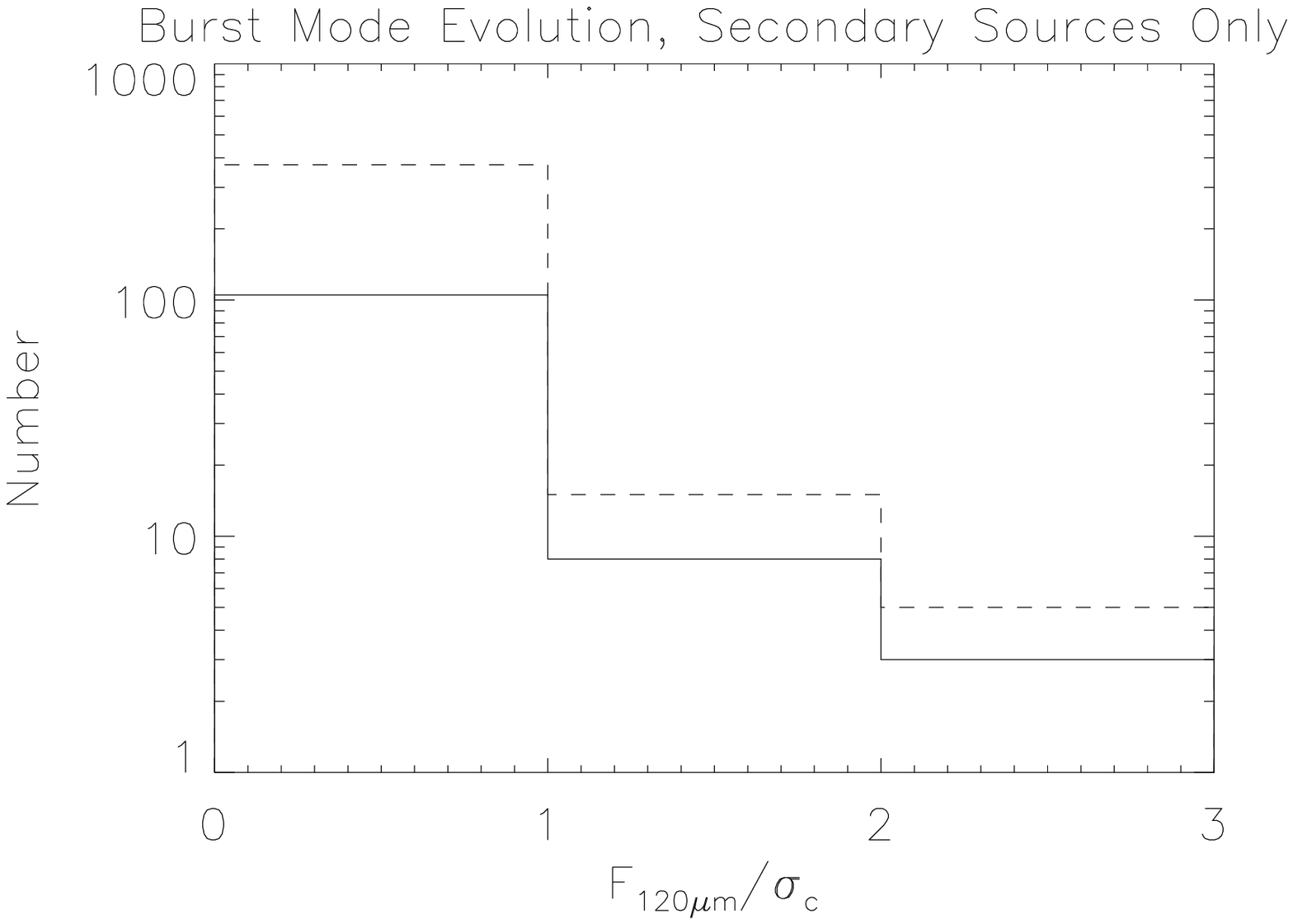, height = 2.5in}
  \caption{Plots of the 120 $\mu$m continuum flux distribution (with
  flux measured as a fraction of the 120 $\mu$m continuum confusion
  limit, $\sigma_{c}(\lambda = 120\mu m) = 4.3$ mJy) for both the
  total input secondary sources (dashed) and secondary sources with
  accurately determined redshifts (solid).  Shown in the left and
  right hand panels are the results for the bright-end and burst mode
  evolution models respectively. Results are for 100 SAFARI FoVs.}
  \label{fig:flux2}
\end{figure*}

Employing a 120 $\mu$m cutoff of 0.342 mJy, we find that we are able
to determine accurate redshifts for $\sim$38(29)\% of secondary
sources with $S_{120\mu m} \geq 0.342$ mJy (see
figure~\ref{fig:flux2}), with 11(18)\% of all redshifts output by the
algorithm under the same constraints being inaccurate.

\section{Discussion}
\label{sec:discussion}
We find that using the PCC method as described in this work, we can
recover redshifts for sources as much as 10 times below the
traditional continuum confusion limit.  We also find we are able to
successfully retrieve redshifts from line confused spectra caused by
multiple sources separated by smaller scales than the size of the beam
on the sky.  However, using the burst mode and bright-end evolutionary
models of \citet{pearson-05}, \citet{pearson-07} and
\citet{pearson-k-09}, we find no spatial bins containing more than two
sources with $S_{120 \mu m} \geq 0.342$ mJy and therefore we do not
test whether or not our method would be able to disentangle more
sources than this.

\begin{figure*}[h!tbp]
  \centering
  \epsfig{figure=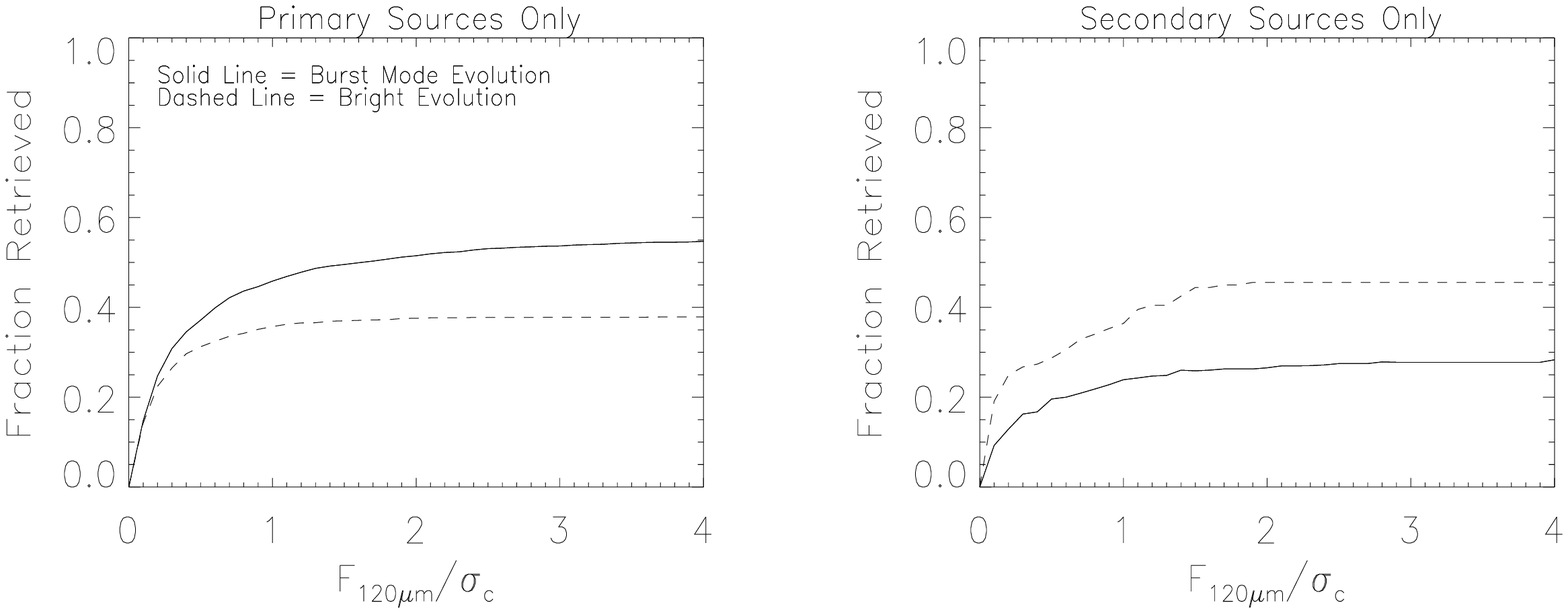, height = 2.5in}
  \caption{Plots of the cumulative fractional recovery of sources
(fraction of sources accurately recovered with $S_{120\mu m}$ less
than that defined by the x-axis) with increasing 120 $\mu$m flux,
measured as a fraction of the traditional 120 $\mu$m continuum
confusion limit, $\sigma_{c}(\lambda = 120\mu m) = 4.3$ mJy.  The left
panel shows our results for primary sources and the right panel shows
our results for secondary sources.  The results for burst mode
evolution are plotted as a solid line, and for bright-end evolution as
a dashed line.}
  \label{fig:recovery}
\end{figure*}

Shown in figure~\ref{fig:recovery} is the cumulative recovery
(fraction of sources accurately recovered with $S_{120\mu m}$ less
than that defined by the x-axis) efficiency with increasing flux.  All
source populations (the primary and secondary sources of both
bright-end and burst mode evolution) have a strongly increasing
cumulative recovery fraction at low fluxes which then levels off at
fluxes higher than the confusion limit.  This is because the bulk of
their populations lie at fluxes fainter than the confusion limit.

The bright-end evolution model has a larger fraction of low flux
sources, whereas the burst mode evolution model has a similar
population of low flux sources, but also a larger number of brighter
sources.  The efficiency of our method falls off with decreasing flux
(as illustrated in figure~\ref{fig:flux1}), therefore we retrieve the
redshifts for the burst mode evolution model more efficiently than for
the bright-end evolution model.

At higher fluxes the redshift determination efficiency for all
populations goes to 100\%, however at these higher fluxes there are
relatively fewer sources.  Even at 10\% of the confusion limit, we are
still retrieving redshifts for $\sim$10\% of sources, and given that
the number of sources at these fluxes is so large we are gaining
information about a significantly larger number of galaxies than would
be the case if we were confusion limited.

The percentage of retrieved redshifts which are in error is higher for
secondary sources than for the primary sources.  This is to be
expected because by definition the secondary sources are fainter than
the primary.  We are thus more likely to confuse noise with genuine
emission lines.  Another difficulty encountered when attempting to
retrieve secondary redshifts is that we are unable to set a continuum
flux limit to consider the source to be viable.  Therefore we are most
often attempting to retrieve redshifts from either very low flux or
non-existent secondary galaxies, significantly increasing the
likelihood of retrieving an erroneous redshift.  This problem is
lessened somewhat by the introduction of more stringent criteria in
other areas when attempting to retrieve secondary redshifts, but it
remains the cause of a large percentage of the inaccurate secondary
redshift recoveries.

We find that by employing a cross correlation method for redshift
determination, where we are only considering a very discrete set of
redshift possibilities, we are able to reduce the $S_{120\mu m}$
cutoff value without drastically decreasing the efficiency of the
method and thus are able to dig deeper into the instrumental noise.
Employing the full MIR and FIR emission lines version of our method we
find that we can retrieve accurate redshifts for a total of 38(52)\%
of all (i.e. both the primary and secondary sources) sources with
$S_{120\mu m} \geq 0.342$ mJy for bright-end and burst mode evolution
respectively.  Their respective frequency of occurrence of inaccurate
redshifts as a percentage of all redshifts output by the algorithm
under the same constraints is 15(10)\%.

A limitation of this method is the possibility of sources with
anomalous line strengths which can, for example, result in a
miss-identification of the [CII] line.  The possibility of such
occurrences has not been taken in to account here.  Further
improvement to our PCC method would aim to decrease the total number
of erroneous redshifts output for reasons such as this.

Additional spectral features that can be used for finding redshift are
those associated with the PAHs (Polycyclic Aromatic Hydrocarbons).
These features are much broader than the emission lines we have been
using, thus observations could be made at lower spectral resolution,
with a corresponding increase in instrument sensitivity.  In this work
we have modeled angular resolution as being constant over the full
SAFARI band.  If we were to employ a more realistic model for angular
resolution we may observe sources which are clustered in a single
spatial bin in our current 8'' binning, as being separable (visible in
different spatial bins) at shorter wavelengths.  This could
significantly decrease the fraction of erroneous redshifts output for
secondary sources.

As discussed earlier, the sources from the bright-end evolution model
are mostly at low fluxes whereas the burst mode has in addition a
smaller population of high flux sources.  This accounts for the
difference in our recovery rates for the two models.  It is therefore
important to further test the efficacy of our technique, and of this
deep observing mode of SAFARI spectral imaging on a number of
different evolutionary models.

Using a Kolmogorov-Smirnov test we are able to determine whether a set
of recovered redshifts are from different parent populations.  Thus by
comparing redshift distributions recovered from datacubes of different
sizes we can determine what area of sky is required for SAFARI to be
able to reliably distinguish between different evolutionary models.
We compare the maximum deviation between the cumulative distributions
\citep{wall-96} of the redshifts output from our PCC method for skies
populated with the burst mode evolution model and with the bright-end
evolution model.  We perform this comparison for increasing numbers of
SPICA FoVs.  Using our PCC redshift determination method we find we
are able to reliably distinguish (probability that the sources from
the two models are drawn from the same distribution, $P=$0.01\%)
between the bright-end and burst mode evolution models with a sky
survey area equal to 8 SAFARI FoVs, each of 10 hrs integration time.

\section{Conclusions}
We have found that our PCC redshift determination method is capable of
resolving sources (i.e. determining a unique redshift) more than an
order of magnitude fainter than the traditional continuum confusion
limit, however the efficacy of our method is higher for brighter
sources.  In this work we have used the PCC method on models based
around the SAFARI instrument for SPICA, however the same technique
could be used on any sensitive imaging spectrometer.  The bright-end
and burst mode evolution models include sources up to redshifts of
$z\sim$4 and 5 respectively.  At these redshifts we are still able to
determine redshifts for sources using the PCC method.  We have not yet
tested to see at which redshift the PCC method begins to fail, this
may be the subject of future work.

The evolutionary models we have investigated in this work have the
bulk of their populations at fluxes fainter than the confusion limit.
By employing our PCC method we are therefore greatly increasing the
number of galaxies that we are able to uniquely identify in any single
observation.  This presents us with a better statistical sample with
which to compare observed source counts and redshift distributions
with those presented in different evolutionary models.  We therefore
have the potential to reliably distinguish different evolutionary
models and our observations with much smaller area surveys, and
therefore within shorter observing times.

Future work should include the following: 1) Use of the PAH features
to identify sources and determine their redshifts.  This will allow us
to decrease our spectral resolution, thus increasing instrument
sensitivity.  2) Investigation of the viability of taking into account
sources with atypical line strengths.  3) More realistic angular
resolution modeling where spatial resolution varies across the
waveband.  4) Investigation of the efficiency of the method when
implemented on a wider range of evolutionary models.  5) A more
quantitative analysis of where our ability to retrieve redshifts from
single and combined spectra begins to break down is being conducted.
6) It should also be noted that since the work described in this paper
was conducted the technical specifications of SAFARI have changed
somewhat (eg. waveband, sensitivity), therefore the results should be
re-checked with more up to date modeling of the SAFARI instrument.
These topics are currently being investigated.

\begin{flushleft}
{\bf Acknowledgements}
\end{flushleft}
We would like to thank the anonymous referee for excellent comments
and suggestions which markedly improved the clarity of the paper. GR
would like to acknowledge an STFC postgraduate studentship, KGI
funding from RCUK and DC was funded in part by STFC.

\end{document}